\documentclass[11pt]{report}

\usepackage{amsfonts}
\usepackage{amsmath}
\usepackage{float}
\usepackage{mathrsfs}
\usepackage{graphicx}

\begin{document}

\begin{center}

\textbf{\large Multifractal analysis of three-dimensional grayscale images: \\
Estimation of generalized fractal dimension \\
and singularity spectrum}

\vspace{0.75cm}

\normalsize Lorenzo Milazzo

\vspace{0.2cm}

\noindent {\small \emph{Cambridge, UK \\ [-0.2ex]
\hspace*{0.05cm} email: lorenzo.milazzo@physics.org}}

\end{center}

\vspace{1.5cm}

\begin{center}

\textbf{Abstract}

\end{center}

\noindent A multifractal analysis is performed on a three-dimensional
grayscale image associated with a complex system. First, a procedure for
generating 3D synthetic images (2D image stacks) of a complex structure
exhibiting multifractal behaviour is described. Then, in order to characterize
the 3D system, the theoretical calculation of the generalized fractal
dimension~$D_{q}$ and two different approaches for evaluating the singularity
spectrum~$f(\alpha)$ are presented.

\vspace{4.0cm}

\noindent \textbf{1. Introduction}

\vspace{0.5cm}

\noindent Geometric (mono)fractals are self-similar sets of
points. Multifractals (or multifractal measures) are self-similar measures
defined on specific set of points [Baveye et al. 2008; Evertsz et al. 1992;
  Falconer 2003]. In general, the former can be generated by additive
processes and the latter by multiplicative cascade of random processes. \\
The \emph{generalized fractal dimension}~$D_{q}$, which is closely related to
the \emph{R\'enyi entropy} [R\'enyi 1961], provides a direct measurement of
the fractal properties of an object -- several values of the \emph{momentum
  order} $q$ correspond to well-known generalized dimensions, such as the
\emph{capacity dimension} (\emph{box-counting dimension})~$D_{0}$, the
\emph{information dimension}~$D_{1}$, and the \emph{correlation
  dimension}~$D_{2}$. The \emph{singularity spectrum}~$f(\alpha)$ provides
information about the scaling properties of the structure [Halsey et al. 1986,
  Hentschel et al. 1983; Lowen et al. 2005; Theiler 1990].

\vspace{0.25cm}

\noindent Nowadays, several techniques can be used to obtain representations
of complex structures through 2D (3D) images. Typically, a 3D image is
constituted by a stack of 2D images -- examples of 3D images are the computed
tomography (CT) images.

\vspace{0.25cm}

\noindent The aim of the present paper is to \emph{a)}~extend a procedure for
generating 2D multifractal lattice (Meakin method) to the 3D case;
\emph{b)}~present the theoretical calculation of the generalized fractal
dimension $D_{q}$ for a 3D multifractal structure generated by a random
multiplicative process; and \emph{c)}~compare two different approaches for
evaluating the singularity spectrum $f(\alpha)$ in the case of 3D images of a
complex structure exhibiting multifractal behaviour.

\vspace{1.7cm}

\noindent \textbf{2. Generation of 3D multifractal lattices: Meakin method}

\vspace{0.5cm}

\noindent In order to investigate the properties of random walks on
multifractal lattices, in 1987 P. Meakin devised a procedure for generating
two-dimensional multifractal lattices by using a \emph{random multiplicative
  process} [Meakin 1987; Stanley et al. 1988]. \\
Recently, new variants of the generator have been proposed [H. Zhou et
  al. 2011]. For this study, we refer to the original algorithm. \\ 
This method is based on an iterative algorithm. Starting with a 2D lattice
containing $2 \times 2$ sites, a 2D lattice with $2^{k} \times 2^{k}$ sites
is generated after $k$ iterations. In the limit $k \to \infty$, the procedure
defines a multifractal measure on the 2D space which can be described in
terms of a countinous spectrum of singularities of type $\alpha$, each
supported on a fractal subset with a fractal dimensionality $f(\alpha)$. \\
To extend the Meakin method to the 3D case, let us consider a finite volume of
linear size $L=1$ and, associated with it, a \emph{3D lattice} containing $2
\times 2 \times 2 = 8$ \emph{boxes} of linear size $l=1/2$. In the first stage
of construction, eight numbers $p_{i}$ are chosen, and randomly associated
with the boxes of the lattice: $p_{1}=p_{2}=p_{3}=p_{4}=\dot{p}$, \,
$p_{5}=p_{6}=p_{7}=p_{8}=\ddot{p}$. Each of these numbers may be regarded as a
\emph{probability}~-- for example, the probability that a specific box is
filled. \\ 
Then, an iterative algorithm is implemented (e.g. by using a \emph{recursive}
function). At each iteration $k$, the lattice is divided in
$N(l) = 2^{k} \times 2^{k} \times 2^{k} = 2^{3k}$ boxes of linear size
$l=2^{-k}$. \\
For the \emph{i}th box, the probability to be filled is:

\begin{equation}
p_{i}(k) = \prod_{j=0}^{k} p_{j} \label{PRMPEq}
\end{equation}

\noindent where $p_{j}=\dot{p}$ or $p_{j}=\ddot{p}$. As a result of this
multiplicative process, there are $(k+1)$ sets of boxes sharing the same
probability value. For each $j$ between 0 and $k$, $2^{2k} \binom{k}{j}$ boxes
of linear size $l=2^{-k}$ and volume $v=2^{-3k}$ have the same probability to
be filled:

\begin{equation}
p_{i} = \dot{p}^{j}\ddot{p}^{k-j}
\end{equation}

\vspace{0.3cm}

\noindent each box having the \emph{mass} equal to:

\begin{equation}
m_{i} = p_{i} \, v = \dot{p}^{j}\ddot{p}^{k-j} \, 2^{-3k}
\end{equation}

\vspace{0.25cm}

\noindent The \emph{total mass} of the system at each iteration $k$ is:

\begin{equation}
m_{T} = \sum_{i=1}^{N} m_{i} = \sum_{j=0}^{k} \left( 2^{2k} \binom{k}{j}
\dot{p}^{j}\ddot{p}^{k-j} \right) 2^{-3k} = 2^{-k} \sum_{j=0}^{k} \binom{k}{j}
\dot{p}^{j}\ddot{p}^{k-j} = 2^{-k} (\dot{p} + \ddot{p})^{k}
\end{equation}

\vspace{0.35cm}

\noindent Hence, for each box $i$ belonging to the set $j$ made of similar
boxes, the \emph{mass probability} (or \emph{mass fraction}) is:

\begin{equation}
p_{i}(l) = \frac{m_{i}}{m_{T}}
        = \frac{\dot{p}^{j}\ddot{p}^{k-j}}{2^{2k} (\dot{p} + \ddot{p})^{k}} \label{MassP1Eq}
\end{equation}

\vspace{1.0cm}

\noindent \emph{Generation of synthetic images} -- The Meakin method has been
used to generate \emph{synthetic images} of 2D complex structures exhibiting
multifractal behaviour [Perrier et al. 2006; H. Zhou et al. 2011]. \\
In this context, the numbers $p_{i}$ generated by using a random
multiplicative process represent \emph{pixel values}. If the images are in
\emph{grayscale} format, the pixel values lay in the range [0,255] and can be
considered as measures of \emph{mass} (0 = black = no mass; 255 = white =
mass). In order to generate a 2D (3D) synthetic image in grayscale format,
the following initial values for the iterative algorithm associated with
Eq.~(\ref{PRMPEq}) are chosen: $\dot{p}=255$ and $\ddot{p}=255/2$. Note that,
at the end of calculations, a rounding procedure has to be applied to ensure
that all pixel values are integer numbers.

\vspace{1.7cm}

\noindent \textbf{3. Multifractal Analysis: Theory}

\vspace{0.5cm}

\noindent

\noindent The \emph{generalized fractal dimension} is defined by:

\begin{equation}
D_{q} = \lim_{l \to 0} \frac{1}{q-1} \frac{\ln{\sum_{i=1}^{N(l)} p_{i}^{q}(l)}}{\ln{l}} \label{DqEq}
\end{equation}

\vspace{0.25cm}

\noindent where $p_{i}(l)$ is the integrated measure (mass probability)
associated with the \emph{i-th} box, $q$ is the momentum order, and $N(l)$ is
the number of boxes of linear size $l$. \\
For $q=1$ (information dimension), the Eq.~(\ref{DqEq}) can be approximated by:

\begin{equation}
D_{1} = \lim_{l \to 0} \frac{\sum_{i=1}^{N(l)} p_{i}(l) \ln{p_{i}(l)}}{\ln{l}}
\end{equation}

\vspace{0.25cm}

\noindent In the case of 3D multifractal structures generated by a random
multiplicative process, the generalized fractal dimension has the following
expressions for $q \ne 1$ and $q = 1$ (see \emph{Appendix~A} for detailed
calculations):

\begin{eqnarray}
D_{q} & = & \frac{ \ln{\left( \displaystyle\frac{\dot{p}^{q} +
      \ddot{p}^{q}}{2^{2(q-1)} (\dot{p} + \ddot{p})^{q}} \right)}} {(1-q)
  \ln{2}} \label{DqRMP1Eq} \\ [0.25cm]
D_{1} & = & \frac{\ln{(2^{2} (\dot{p} + \ddot{p}))} - \left(\dot{p} \,
    \ln{(\dot{p})} + \ddot{p} \, \ln{(\ddot{p})} \right) (\dot{p} +
    \ddot{p})^{-1}}{\ln{2}} \label{DqRMP2Eq}
\end{eqnarray}

\vspace{0.25cm}

\noindent Once $D_{q}$ is known, the \emph{singularity spectrum} $f(\alpha)$ can be evaluated
via a Legendre transformation:

\begin{equation}
f(\alpha(q)) = q \alpha(q) - \tau(q), \hspace*{1.0cm} \alpha(q) = \frac{d\tau(q)}{dq}
\end{equation}

\noindent where $\tau(q) = (q-1)D_{q}$ [Halsey et al. 1986].

\vspace{0.25cm}

\noindent The singularity spectrum can also be directly (without knowing
$D_{q}$) evaluated by using the method proposed by Chhabra \emph{et al.} [1989].
The first step of this approach consists of defining a family of normalized
measures $\mu(q)$:

\begin{equation}
\mu_{i}(q,l) = \frac{[p_{i}(l)]^{q}}{\sum_{j=1}^{N(l)}[p_{j}(l)]^{q}}
\end{equation}

\vspace{0.25cm}

\noindent For each box~\emph{i}, the normalized measure $\mu_{i}(q,l)$ depends
on the order of the statistical moment and on the box size and it takes values
in the range [0,1] for any value of $q$. \\
Then, the two functions $f(q)$ and $\alpha(q)$ are evaluated:

\begin{eqnarray}
f(q)      & = & \lim_{l \to 0} \frac{\sum_{i=1}^{N(l)}\mu_{i}(q,l) \ln \mu_{i}(q,l)}{\ln l} \\
\alpha(q) & = & \lim_{l \to 0} \frac{\sum_{i=1}^{N(l)}\mu_{i}(q,l) \ln p_{i}(l)}{\ln l}
\end{eqnarray}

\vspace{0.25cm}

\noindent For each $q$, values of $f(q)$ and $\alpha(q)$ are obtained from the
slope of plots of $\sum_{i=1}^{N(l)}\mu_{i}(q,l) \ln \mu_{i}(q,l)$ versus
$(\ln l)$ and $\sum_{i=1}^{N(l)}\mu_{i}(q,l) \ln p_{i}(l)$ versus $(\ln l)$
over the entire range of box size values under consideration. Finally, the
singularity spectrum $f(\alpha)$ is constructed from these two data sets.

\vspace{1.7cm}

\noindent \textbf{4. Results and Discussion}

\vspace{0.5cm}

\noindent For this study, a stack of 128 images with a resolution of
($128\times128$) pixels is generated by using the Meakin method
(Eq.~(\ref{PRMPEq}), number of iterations $k=7$). The image stack represents a
3D image of a finite volume of linear size $L=128$ [pixels]. For this complex
structure, we evaluate the singularity spectrum $f(\alpha)$ by using the two
different methods previously introduced; in order to compare the \emph{indirect}
and \emph{direct} approaches, we adopt the same set of 21 values of moment
orders: $q~=~\{-5.0, -4.5, -4.0, \dots, 4.0, 4.5, 5.0\}$.

\vspace{0.25cm}

\noindent To indirectly evaluate $f(\alpha)$, we calculate $D_{q}$
first. Since the synthetic images have been generated by using a random
multiplicative process, the generalized fractal dimension can be derived from
Eq.~(\ref{DqRMP1Eq}) and (\ref{DqRMP2Eq}) with $\dot{p}=255$ and
$\ddot{p}=255/2$. Fig. 1 shows $D_{q}$ as function of the order moment. Then,
we derive the singularity strength and the Hausdorff dimension of normalized
measures as function of $q$ (Fig. 2) and, finally, the singularity spectrum
(Fig. 3). As expected, the capacity dimension ($q=0$) is ~$D_{0}=3$ and the
curve $f(\alpha)$ is convex with a single maximum at $q=0$ ($\alpha=3.085$)
and with infinite slope at $q=\mp\infty$ [Halsey et al. 1986]. \\
Note that, for a multifractal system, the variation of generalized fractal
dimension with the order moment quantifies the nonuniformity. The maximum
value of the $D_{q}$ is associated with the least-dense points on the fractal
and the minimum value corresponds to the most-dense points [Theiler 1990].

\vspace{0.25cm}

\noindent In order to perform multifractal analysis of 2D (3D) grayscale
images and, in particular, to directly evaluate the singularity spectrum
$f(\alpha)$ by using the Chhabra method, a program -- \texttt{Munari} -- has
been developed [Milazzo 2010]. The application is written in~C++ and, at this
stage, it processes grayscale images in ASCII format. \\
During the initial processing stage, the mass probabilities
(Eq.~(\ref{MassP1Eq})) associated with the box-shaped parts of the images need
to be evaluated. A partitioning procedure could be used to carry out this
calculation -- at each step, a cell (a square box in 2D; a cubic box in 3D) is
divided into sub-cells ($2^{2}$ in 2D, $2^{3}$ in 3D) and $p_{i}(l)$ is
calculated for each of them. However, a different approach is adopted by the
\texttt{Munari} program: an aggregation procedure consisting of grouping
together $2^{2}$ adjoining cells in the 2D case and $2^{3}$ in the 3D
case. This choice is determined by the fact that it is less computationally
intensive -- the higher the number of the initial measures (i.e. the number of
pixels) to be processed, the more efficient the aggregation is in comparison
to the partitioning procedure. Note that the algorithms within \texttt{Munari}
have been designed to be general and, as a result, they are independent from
the resolution of the images and from the values of box sizes and moment
orders used by the Chhabra method. \\
This specific image stack ($128\times128\times128$ pixels) is processed by
using six values of box sizes: $l = \{2, 4, 8, 16, 32, 64\}$. The singularity
spectrum that is obtained after the calculations is identical to the curve
$f(\alpha)$ evaluated by using the indirect method (Fig. 3).

\vspace{1.7cm}

\noindent \textbf{5. Conclusions}

\vspace{0.5cm}

\noindent A multifractal analysis has been carried out on a three-dimensional
grayscale image. The synthetic images were generated by using a random
multiplicative process. \\
First, we have extended a procedure for generating multifractal lattice
(Meakin method) and the theoretical calculation of the generalized fractal
dimension for this kind of structure, to the 3D case. Then, two different methods
(indirect and direct) have been used to evaluate the singularity spectrum for
the complex structure under analysis. They both have given the same
result. Finally, we have presented \texttt{Munari}, a new program for image
processing to carry out multifractal analysis on 2D and 3D systems. This study
is part of the validation tests that have been and will continue to be
implemented to ensure reliability and stability of the application.

\newpage

\noindent \textbf{Appendix A}

\vspace{0.5cm}

\noindent \emph{Theoretical calculation of the generalized fractal dimension
  for a 3D multifractal structure generated by a random multiplicative process}

\vspace{1.0cm}

\noindent The Meakin method is based on an iterative algorithm. Let us
consider a finite volume of linear size $L=1$ and, associated with it, a 3D
lattice containing $2 \times 2 \times 2 = 8$ boxes of linear size $l=1/2$. At
each iteration $k$, the lattice is divided in $N(l) = 2^{k} \times 2^{k}
\times 2^{k} = 2^{3k}$ boxes of linear size $l=2^{-k}$. Each box $i$ belonging
to the set $j$ made of similar boxes has the integrated measure (mass
probability) given by Eq.~(\ref{MassP1Eq}). \\
Thus, the generalized fractal dimension for $q \ne 1$ and $q = 1$ can be
derived as follows:

\vspace*{0.75cm}

\begin{eqnarray}
D_{q} & = & \lim_{l \to 0} \frac{1}{q-1} \frac{\ln{\sum_{i=1}^{N(l)}
    p_{i}^{q}(l)}}{\ln{l}} = \lim_{k \to \infty} \frac{1}{q-1}
\displaystyle\frac{\ln{\sum_{j=0}^{k} 2^{2k} \binom{k}{j} \left(
    \displaystyle\frac{\dot{p}^{j}\ddot{p}^{k-j}}{2^{2k}(\dot{p} +
      \ddot{p})^{k}} \right)^q}}{\ln{2^{-k}}} = \nonumber \\ [0.5cm]
 & = & \lim_{k \to \infty} \frac{1}{q-1} \frac{ \ln{\left(
    \displaystyle\frac{2^{2k}}{(2^{2k} (\dot{p} + \ddot{p})^{k})^{q}}
    \sum_{j=0}^{k} \binom{k}{j} (\dot{p}^{q})^{j}(\ddot{p}^{q})^{k-j}
    \right)}} {\ln{2^{-k}}} = \nonumber \\ [0.25cm]  
 &   & \\ [-0.25cm]
 & = & \lim_{k \to \infty} \frac{1}{q-1} \frac{ \ln{\left( \displaystyle\frac{(\dot{p}^{q}
      + \ddot{p}^{q})^{k}}{2^{2k(q-1)} ((\dot{p} + \ddot{p})^{q})^{k}}
    \right)}} {\ln{2^{-k}}} = \nonumber \\ [0.5cm]
 & = & \lim_{k \to \infty} \frac{1}{q-1} \frac{ k \ln{\left(
    \displaystyle\frac{\dot{p}^{q} + \ddot{p}^{q}}{2^{2(q-1)} (\dot{p} +
      \ddot{p})^{q}} \right)}} {-k \ln{2}} = \nonumber \\ [0.5cm]
 & = & \frac{ \ln{\left( \displaystyle\frac{\dot{p}^{q} +
      \ddot{p}^{q}}{2^{2(q-1)} (\dot{p} + \ddot{p})^{q}} \right)}} {(1-q) \ln{2}} \nonumber
\end{eqnarray}

\vspace{1.0 cm}

\begin{eqnarray}
D_{1} & = & \lim_{l \to 0} \frac{\sum_{i=1}^{N(l)} p_{i}(l)
  \ln{p_{i}(l)}}{\ln{l}} = \lim_{k \to \infty} \frac{\sum_{j=0}^{k} 2^{2k}
  \binom{k}{j} \left( \displaystyle\frac{\dot{p}^{j}\ddot{p}^{k-j}}{2^{2k}
    (\dot{p} + \ddot{p})^{k}}\right) \ln{\left(
    \displaystyle\frac{\dot{p}^{j}\ddot{p}^{k-j}}{2^{2k} (\dot{p} +
      \ddot{p})^{k}}\right)}}{\ln{2^{-k}}} = \nonumber \\ [0.5cm]
 & = & \lim_{k \to \infty} \frac{\sum_{j=0}^{k} \binom{k}{j} \,
  \dot{p}^{j}\ddot{p}^{k-j} \ln{\left(
    \displaystyle\frac{\dot{p}^{j}\ddot{p}^{k-j}}{2^{2k} (\dot{p} +
      \ddot{p})^{k}}\right)}}{-k \, (\dot{p} + \ddot{p})^{k} \ln{2}} = \nonumber \\ [0.5cm] 
 & = & \lim_{k \to \infty} \frac{\sum_{j=0}^{k} \binom{k}{j} \,
  \dot{p}^{j}\ddot{p}^{k-j} \ln{(\dot{p}^{j}\ddot{p}^{k-j})}}{-k \, (\dot{p} +
  \ddot{p})^{k} \ln{2}} - \frac{\sum_{j=0}^{k} \binom{k}{j} \,
  \dot{p}^{j}\ddot{p}^{k-j} \ln{(2^{2} (\dot{p} + \ddot{p}))^{k}}}{-k \,
    (\dot{p} + \ddot{p})^{k} \ln{2}} = \nonumber \\ [0.5cm]
 & = & \lim_{k \to \infty} \frac{\ln{(\dot{p})} \, \sum_{j=0}^{k} j \binom{k}{j} \,
  \dot{p}^{j}\ddot{p}^{k-j} + \ln{(\ddot{p})} \, \sum_{j=0}^{k} (k-j)
  \binom{k}{j} \, \dot{p}^{j}\ddot{p}^{k-j}}{-k \, (\dot{p} +
  \ddot{p})^{k} \ln{2}} + \nonumber \\ [0.5cm]
 &   & + \frac{\ln{(2^{2} (\dot{p} + \ddot{p}))^{k}} \, \sum_{j=0}^{k} \binom{k}{j} \,
  \dot{p}^{j}\ddot{p}^{k-j}}{k \, (\dot{p} + \ddot{p})^{k} \ln{2}} = \nonumber \\ [0.5cm]
 &   & \\ [-0.25cm]
 & = & \lim_{k \to \infty} \frac{\ln{(\dot{p})} \,\, k\dot{p} \,\, \sum_{j=0}^{k-1} \binom{k-1}{j} \,
  \dot{p}^{j}\ddot{p}^{k-1-j} + \ln{(\ddot{p})} \,\, k\ddot{p} \,\, \sum_{j=0}^{k-1}
  \binom{k-1}{j} \, \dot{p}^{j}\ddot{p}^{k-1-j}}{-k \, (\dot{p} +
  \ddot{p})^{k} \ln{2}} + \nonumber \\ [0.5cm]
 &   & + \frac{\ln{(2^{2} (\dot{p} + \ddot{p}))^{k}} \, \sum_{j=0}^{k} \binom{k}{j} \,
  \dot{p}^{j}\ddot{p}^{k-j}}{k \, (\dot{p} + \ddot{p})^{k} \ln{2}} = \nonumber \\ [0.5cm]
 & = & \lim_{k \to \infty} \frac{k \left(\dot{p} \, \ln{(\dot{p})} \, (\dot{p} +
    \ddot{p})^{k-1} + \ddot{p} \, \ln{(\ddot{p})} \, (\dot{p} +
    \ddot{p})^{k-1} \right)}{-k \, (\dot{p} + \ddot{p})^{k} \ln{2}} + \nonumber \\ [0.5cm]
 &   & + \frac{k \, \ln{(2^{2} (\dot{p} + \ddot{p}))} \, (\dot{p} +
    \ddot{p})^{k}}{k \, (\dot{p} + \ddot{p})^{k} \ln{2}} = \nonumber \\ [0.5cm]
 & = & \frac{ \left(\dot{p} \, \ln{(\dot{p})} + \ddot{p} \, \ln{(\ddot{p})}
    \right) (\dot{p} + \ddot{p})^{-1} }{-\ln{2}} + \frac{\ln{(2^{2} (\dot{p} +
      \ddot{p}))}}{\ln{2}} = \nonumber \\ [0.5cm]
 & = & \frac{\ln{(2^{2} (\dot{p} + \ddot{p}))} - \left(\dot{p} \,
    \ln{(\dot{p})} + \ddot{p} \, \ln{(\ddot{p})} \right) (\dot{p} +
    \ddot{p})^{-1}}{\ln{2}} \nonumber
\end{eqnarray}

\newpage

\noindent \textbf{References}

\vspace{0.5 cm}

\noindent {\small
P. Baveye et al., Introduction to fractal geometry, fragmentation processes
and multifractal measures: Theory and operational aspects of their application
to natural systems, in N. Senesi and K.J. Wilkinson~(Ed.), \emph{Biophysical
  Chemistry of Fractal Structures and Processes in Environmental Systems},
John Wiley \& Sons, Chichester, 11-67 (2008) \\ [1.0ex]
A.B. Chhabra et al., Direct determination of the $f(\alpha)$ singularity spectrum and
its application to fully developed turbulence, \emph{Phys. Rev. A} 40, 5284-5294 (1989) \\ [1.0ex]
C.J.G. Evertsz and B.B. Mandelbrot, Multifractal measures, in H.O. Peitgen et
al.~(Ed.), \emph{Chaos and Fractals: New Frontiers of Science},
Springer-Verlag, New York, Appendix B, 921-953 (1992) \\ [1.0ex]
K.J. Falconer, \emph{Fractal Geometry: Mathematical Foundations and
  Applications}, Wiley, Chichester (2003) \\ [1.0ex]
T.C. Halsey et al., Fractal measures and their singularities: The
characterization of strange sets, \\
\emph{Phys. Rev. A} 33, 1141-1151 (1986) \\ [1.0ex]
H.G.E. Hentschel and I. Procaccia, The infinite number of generalized
dimensions of fractals and strange attractors, \emph{Physica D}, 8, 435-444
(1983) \\ [1.0ex]
S.B. Lowen and M.C. Teich, \emph{Fractal-Based Point Processes},
Wiley-Interscience, New Jersey (2005) \\ [1.0ex]
P. Meakin, Random walks on multifractal lattices, \emph{J. Phys. A:
  Math. Gen.} 20, L771-L777 (1987) \\ [1.0ex]
L. Milazzo, \emph{Munari -- A program for multifractal characterization of
  complex systems}, User Manual (2010) \\ [1.0ex]
E. Perrier, A.M. Tarquis, and A. Dathe, A program for fractal and multifractal
analysis of two-dimensional binary images: Computer algorithms versus
mathematical theory, \emph{Geoderma} 134, 284-294 (2006) \\ [1.0ex]
A. R\'enyi, On measures of information and entropy, \emph{Proc. 4th
Berkeley Sympos. Math. Statist. Probab.}, University of California Press,
Berkeley, 547-561 (1961) \\ [1.0ex]
H.E. Stanley and P. Meakin, Multifractal phenomena in physics and chemistry, \\
\emph{Nature} 335, 405-409 (1988) \\ [1.0ex]
J. Theiler, Estimating fractal dimension, \emph{J. Opt. Soc. Am. A} 7,
1055-1073 (1990) \\ [1.0ex]
H. Zhou et al., Multifractal analyses of grayscale and binary soil thin section
images, \emph{Fractals} 19(3), 299-309 (2011)

}

\newpage

\begin{figure}[!h]
\rotatebox{-90}{
\centering
\includegraphics[scale=0.55]{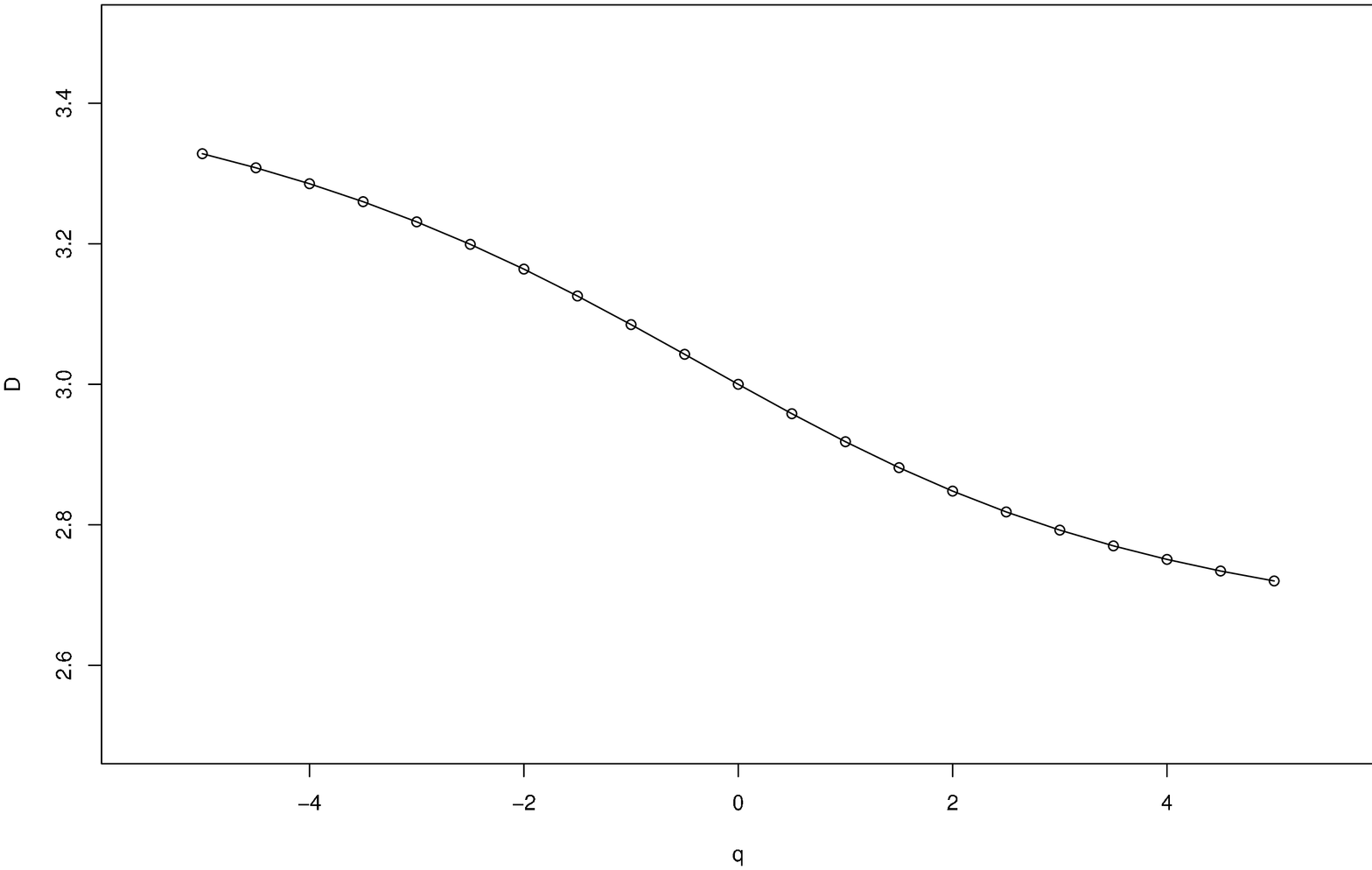}
}
\end{figure}

\vspace{-0.2 cm}

\noindent \textbf{Fig. 1} -- The generalized fractal dimension $D_{q}$ as function of the order moment.

\vspace{2.5 cm}

\begin{figure}[!h]
\rotatebox{-90}{
\centering
\includegraphics[scale=0.55]{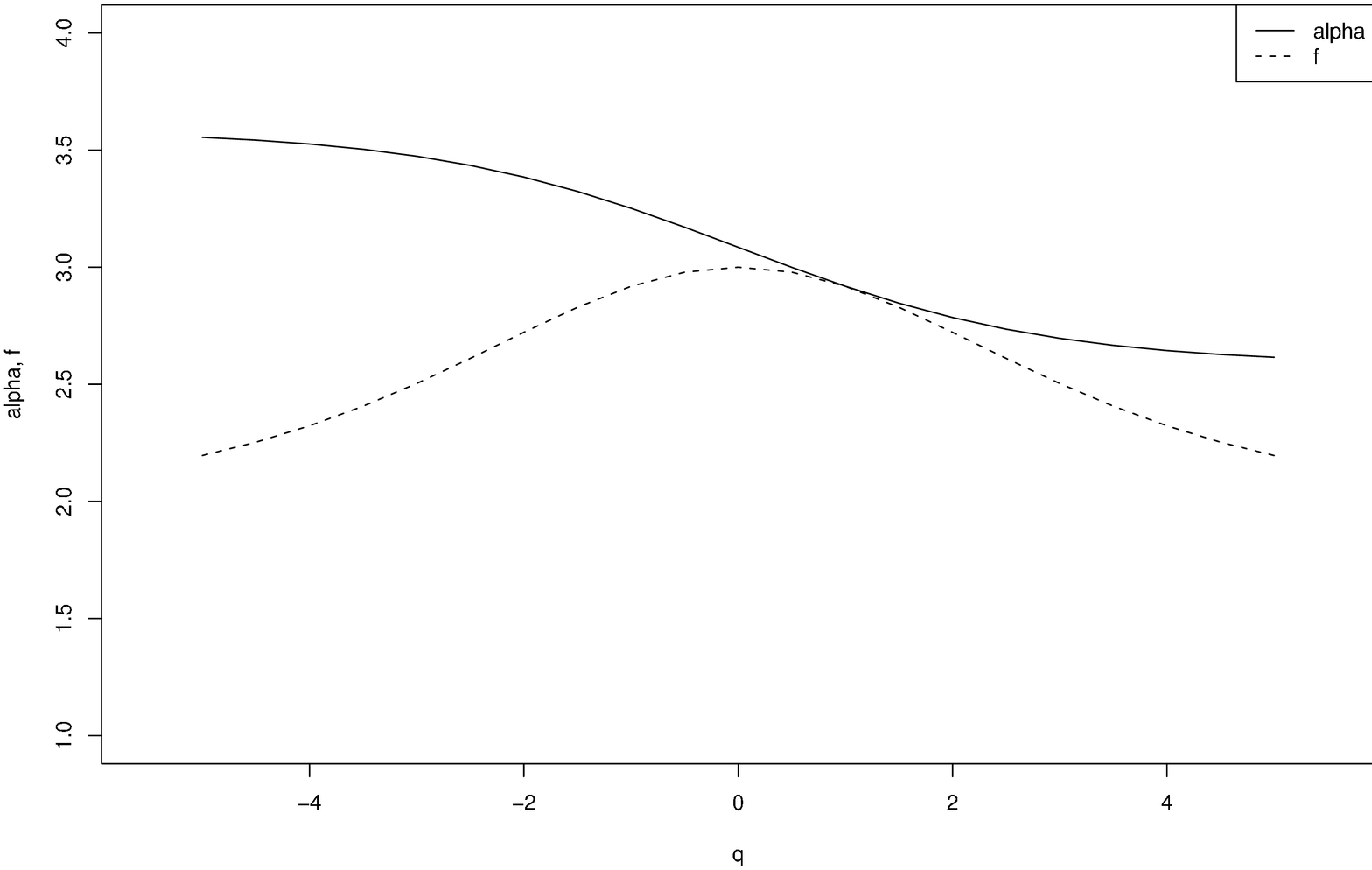}
}
\end{figure}

\vspace{-0.2 cm}

\noindent \textbf{Fig. 2} -- The singularity strength $\alpha(q)$ and the Hausdorff dimension of normalized \\
\hspace*{1.5cm} measures $f(q)$ as function of the order moment.

\newpage

\begin{figure}[!h]
\rotatebox{-90}{
\centering
\includegraphics[scale=0.55]{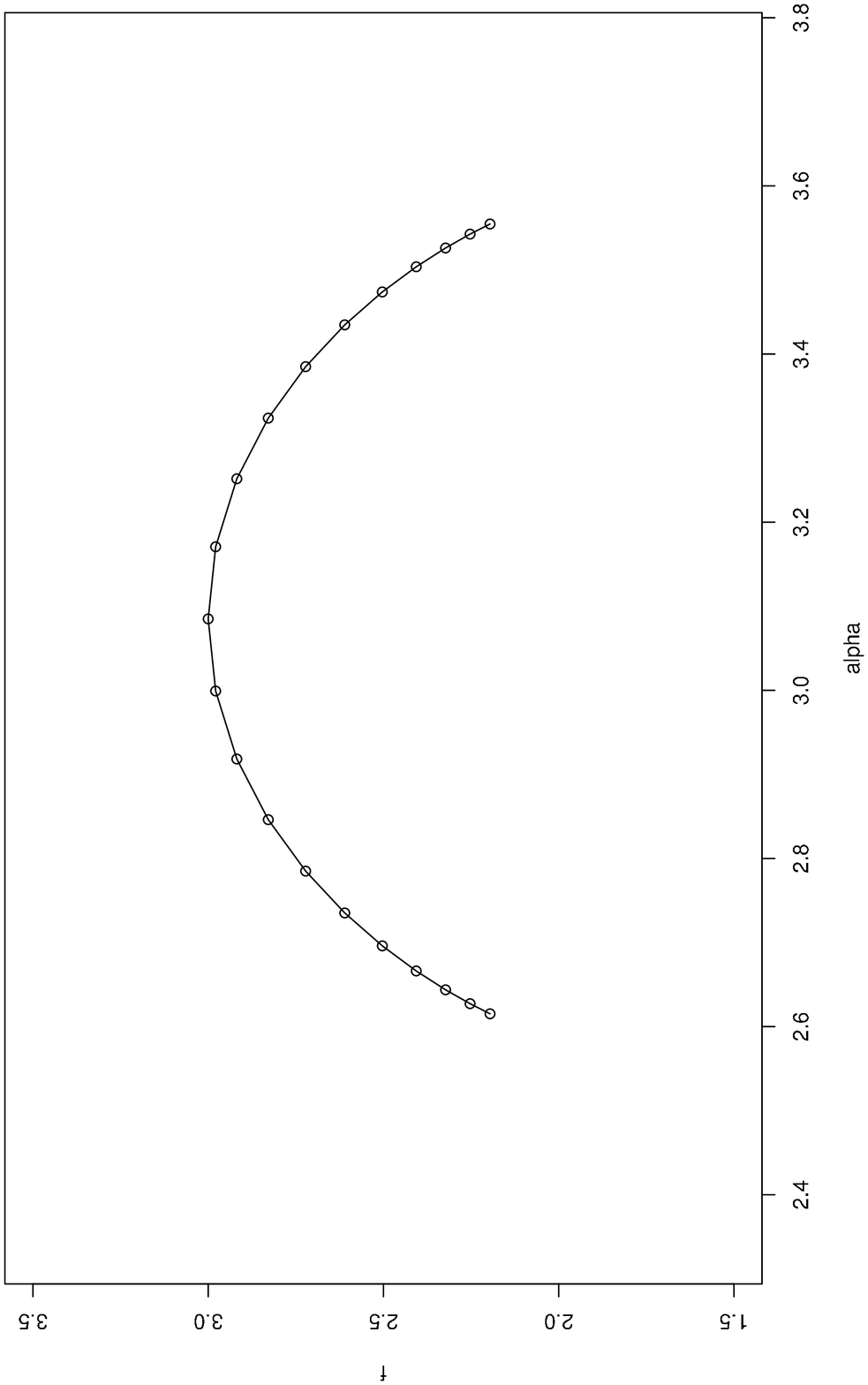}
}
\end{figure}

\noindent \textbf{Fig. 3} -- The singularity spectrums $f(\alpha)$ obtained by
using the \emph{indirect} and \emph{direct} methods. \\
\hspace*{1.5cm} The two curves coincide.

\end{document}